\documentclass[aps,prB,twocolumn,superscriptaddress,floatfix,longbibliography]{revtex4}
\usepackage{amsfonts}
\usepackage{amssymb}
\usepackage{graphicx}
\usepackage{dcolumn}
\usepackage{bm}
\usepackage{amsmath}
\usepackage[colorlinks,linkcolor=magenta,citecolor=blue,urlcolor=blue]{hyperref}
\usepackage{changes}
\begin{document}
	
\preprint{This line only printed with preprint option}

\title{Emergent topological re-entrant phase transition in a generalized quasiperiodic modulated Su-Schrieffer-Heeger model}

\author{Xiao-Ming Wang}
\affiliation {Key Laboratory of Atomic and Subatomic Structure and Quantum Control (Ministry of Education), Guangdong Basic Research Center of Excellence for Structure and Fundamental Interactions of Matter, School of Physics, South China Normal University, Guangzhou 510006, China} 
\affiliation {Guangdong Provincial Key Laboratory of Quantum Engineering and Quantum Materials, Guangdong-Hong Kong Joint Laboratory of Quantum Matter, Frontier Research Institute for Physics, South China Normal University, Guangzhou 510006, China}

\author{Shan-Zhong Li}
\email[Corresponding author: ]{szhongli@m.scnu.edu.cn}
\affiliation {Key Laboratory of Atomic and Subatomic Structure and Quantum Control (Ministry of Education), Guangdong Basic Research Center of Excellence for Structure and Fundamental Interactions of Matter, School of Physics, South China Normal University, Guangzhou 510006, China} 
\affiliation {Guangdong Provincial Key Laboratory of Quantum Engineering and Quantum Materials, Guangdong-Hong Kong Joint Laboratory of Quantum Matter, Frontier Research Institute for Physics, South China Normal University, Guangzhou 510006, China}

\author{Zhi Li}
\email[Corresponding author: ]{lizphys@m.scnu.edu.cn}
\affiliation {Key Laboratory of Atomic and Subatomic Structure and Quantum Control (Ministry of Education), Guangdong Basic Research Center of Excellence for Structure and Fundamental Interactions of Matter, School of Physics, South China Normal University, Guangzhou 510006, China} 
\affiliation {Guangdong Provincial Key Laboratory of Quantum Engineering and Quantum Materials, Guangdong-Hong Kong Joint Laboratory of Quantum Matter, Frontier Research Institute for Physics, South China Normal University, Guangzhou 510006, China}

\date{\today}

\begin{abstract}
We study the topological properties of the one-dimensional generalized quasiperiodic modulated Su-Schrieffer-Heeger model. The results reveal that topological re-entrant phase transition emerges. Through the analysis of a real-space winding number , we divide the emergent topological re-entrant phase transitions into two types. The first is the re-entrant phase transition from the traditional topological insulator phase into the topological Anderson insulator phase, and the second is the re-entrant phenomenon from one topological Anderson insulator phase into another topological Anderson insulator phase. These two types of  re-entrant phase transition correspond to bounded and unbounded cases of quasiperiodic modulation, respectively.  Furthermore, we verify the above topological re-entrant phase transitions by analyzing the Lyapunov exponent and bulk gap. Since Su-Schrieffer-Heeger models have been realized in various artificial systems (such as cold atoms, optical waveguide arrays, ion traps, Rydberg atom arrays, etc.), the two types of topological re-entrant phase transition predicted in this paper are expected to be realized in the near future.
 
\end{abstract}

\maketitle

\section{Introduction}
Topological insulators (TI), as a system with unique transport properties, constitute one of the most important research directions in condensed matter physics and quantum computation~\cite{x1,x2,x3,x4,x5,x6,x7,x8,x10,x13,x14,x15,x16,x17,x18,x19,x20,1,2,3,4,5,6,7,8,9,10}. Previous studies have shown that the non-trivial edge states in topological systems feature good robustness, which ensures that the edge current can still maintain its original state in the case of weak disorder or defects. However, when disorder becomes very strong, the topological system will undergo Anderson phase transition~\cite{03}. In other words, in the case of strong disorder, the topological edge current will be destroyed to make the topological system a gapless Anderson insulator, where the corresponding bulk states show the characteristics of localized states. In addition, recent studies have revealed that moderate disorder can achieve a transition from a trivial phase to a non-trivial phase, and this topological system induced by disorder is called topological Anderson insulator (TAI)~\cite{001,traceb0,04,05,06,07,f6,f7,f9,f10,h01,h02,h03,h04,h05,h06,h07,h08,h1,h09}.

In recent years, TAIs and its related fields have been greatly developed, and a series of milestone achievements have been scored. Theoretically, in addition to the standard TAI, the following systems have also been predicted: $\mathbb{Z}_{2}$ topological Anderson insulators~\cite{e1}, Topological Anderson amorphous insulator~\cite{e2}, Higher-order Topological Anderson Insulators~\cite{e3,h2}, Topological inverse Anderson insulator~\cite{e4}, etc~\cite{e5,e6,e7}. Besides, with the increasing popularity of non-Hermitian research~\cite{nh00,nh001,nh1,nh002,nh01,nh02,nh03,nh04,nh05}, the study on TAI has gradually extended to non-Hermitian systems~\cite{z1,z02,z03,z04}. Experimentally, TAI has been realized in a variety of artificial systems, including ultra-cold atoms~\cite{z2,z3,zz0}, photonic/phononic system~\cite{z4,z5,z6}, superconducting system~\cite{002}, and electric circuits~\cite{z7}%Phys. Rev. Lett. 126, 146802 (2021)
, etc.

Re-entrant phase transition (REPT), on the other hand, refers to the process in which the system starts from a phase and returns to the same phase by monotonically manipulating a certain parameter~\cite{z8}.  Recently, the REPT of localized phase has been discovered in Aubry-Andr\'{e} model~\cite{a0,a1} and SSH model~\cite{519,512}, etc~\cite{520,521}. Moreover, topological REPT has also been reported recently~\cite{tr1}.  So far, although there are many researches on TI and TAI, few work has been done on topological REPT in TI and TAI~\cite{tr2}. This paper is devoted to the study of REPT phenomena in TI and TAI system.
%%%%%%
%%%%%%
\begin{figure}[htbp]
\centering
\includegraphics[width=8.3cm]
{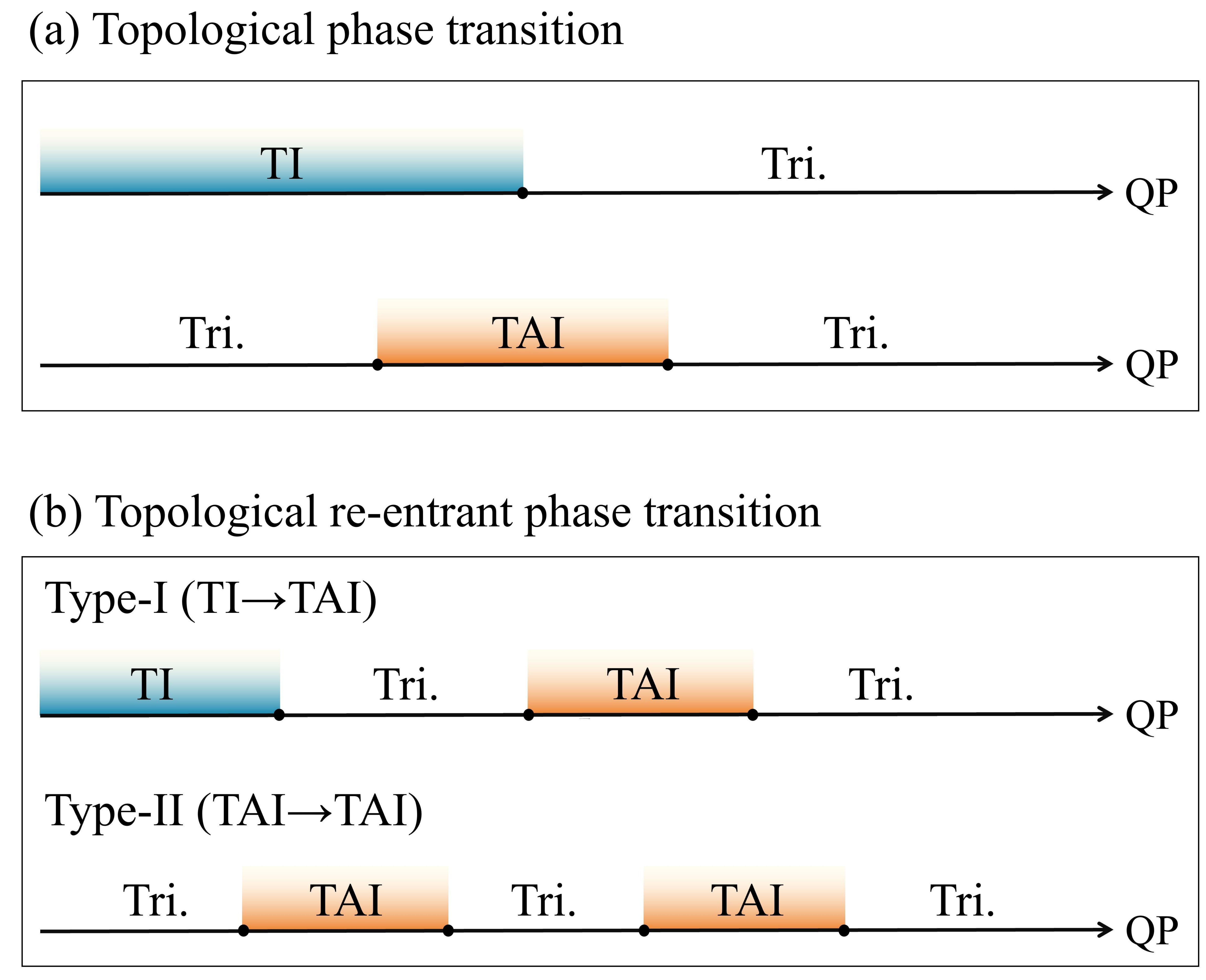}
\caption{Schematic diagram of quasiperiodic induced traditional topological phase transitions and topological REPT.}
\label{F1}
 \end{figure}
%%%%%%
%%%%%%

There are two main findings in this paper. First, the generalized quasiperiodic modulation can induce topological REPTs in one-dimensional Su-Schrieffer-Heeger (SSH) model. Second, the emerged REPTs can be divided into two classes. In concrete terms, when the bounded (unbounded) structure quasiperiodic modulation has been selected, REPTs of TI$\rightarrow$TAI (TAI$\rightarrow$ TAI) will emerge~[see Fig.~\ref{F1}].

The rest of this paper is organized as follows. In sec.~\ref{S2}, we briefly introduce the one-dimensional SSH model with generalized quasiperiodic modulation. In sec.~\ref{S3}, under the condition of the bounded case, we discuss the first type of REPT by computing winding number, energy gap and Lyapunov exponent. In sec.~\ref{S4}, we discuss the unbounded case. The main results of this paper are summarized in sec.~\ref{S6}.

\section{MODEL and Key quantities}\label{S2}
Let's start at the generalized quasiperiodic modulated SSH model. The corresponding Hamiltonian reads
%%%
%%%
\begin{equation}
\begin{aligned}
H=\sum_{n=1}^N\left(t_1^{'} a_n^{\dagger} b_n+t_2 a_{n+1}^{\dagger} b_n+\text { H.c. }\right),
\end{aligned}\label{Hamiltonian}
\end{equation}
%%%
%%%
where $a_n$ ($b_n$) is the annihilation operator for the sublattice A (B) on n-th primitive cell. $N$ is the total number of primitive cells. $t_1^{'}$ and $t_2$ denote the intracell and the intercell hopping strength, respectively. Here, we consider applying a generalized quasiperiodic modulation on the intracell hopping term~\cite{san0}, i.e., 
%%%
%%%
\begin{equation}\label{eq2}
{t_1^{'}}=t_1+\frac{\lambda \cos (2 \pi \alpha n+\theta)}{1-b \cos (2 \pi \alpha n+\theta)},
\end{equation}
%%%
%%%
where $t_1$ is the intracell hopping strength, $\lambda$ is the strength of quasiperiodic modulation, $b$ is the structure factor, which is the key parameter to control the quasiperiodic modulation bounded ($\left|b\right|<1$) or unbounded ($\left|b\right|\geq1$). $\theta$ is an additional phase shift, and $\alpha$ is an irrational number. When $\lambda=0$, the Hamiltonian Eq.~\eqref{Hamiltonian} reduces to the standard SSH model~\cite{ssh1}. The system represents a trivial (non-trivial) topology under the condition of $t_1<t_2$ $(t_1>t_2)$. Without loss of generality, we set $t_2=1$ as the energy unit. We take $\alpha=(\sqrt{5}-1)/2$ and $\theta=0$. In the numerical calculation, we choose the system size $N=L/2=610$ with $L$ being the total lattice number. Such size is large enough for self-averaging and one can safely ignore the finite size effect (see Appendix~\ref{Sec.7} for details). 

In this paper, since we are concerned with topological properties, we mainly discuss three quantities related to topological properties.

The first one is the winding number, which can well reflect the system's topological properties. Since the generalized quasiperiodic modulation breaks the translational symmetry, generally speaking, one can use the real-space winding number as the indicator to characterize the topological properties of a quasiperiodic topological system~\cite{san1}, which can be defined as
%%%
%%%
\begin{equation}
\nu=\frac{1}{L^{'}} \operatorname{Tr}^{'}(\Gamma \mathrm{Q}[\mathrm{Q}, \mathrm{X}]).
\label{eq3}
\end{equation}
%%%
%%%
where $\text{Q}={\textstyle\sum_{j=1}^{N}}(\left|j\right\rangle \left\langle j\right|-\left|\tilde{j}\right\rangle\left\langle\tilde{j}\right|)$ is the corresponding open-boundary matrix and $\left|\tilde{j}\right\rangle=\Gamma^{-1}\left|j\right\rangle$. One can directly obtain the matrix $\text{Q}$ by solving the eigenequations $H\left|j\right\rangle=E_{j}\left|j\right\rangle$, where $E_{j}$ and $\left|j\right\rangle$ correspond to the eigenenergies and eigenstates, respectively. $\Gamma=I_{N}\otimes\sigma_{z}$ denotes the chiral symmetry operator with the identity matrix $I_{N}$ and the Pauli matrix $\sigma_{z}$. The key operator
%%%
%%%
\begin{equation}
\text{X}=\begin{pmatrix}
  1  &0  &0  &0  &0  &\cdots &0  &0\\
  0  &1  &0  &0  &0  &\cdots &0  &0 \\
  0  &0  &2  &0  &0  &\cdots &0  &0\\
  0  &0  &0  &2  &0  &\cdots &0  &0\\
  \vdots & \vdots & \vdots & \vdots & \ddots   &\cdots & \vdots &\vdots\\
  0  &0  &0  &0  &0  &\cdots &N  &0\\
  0  &0  &0  &0  &0  &\cdots &0  &N\\
\end{pmatrix}
\label{eq3}
\end{equation} 
%%%
%%%
is the coordinate operator. The symbol $\operatorname{Tr}^{'}$ means the trace over the middle interval of the full lattice with the length $L^{'}=L/2=N$ (see Appendix~\ref{Sec.7} for details). For example, under the condition of the primitive cells' number $N=6$. Since a primitive cell contains two sublattices, the total number of sites is $12$. The corresponding matrix in Eq.~\eqref{eq3} is a $12\times12$ matrix. Then, $\operatorname{Tr}^{\prime}$ represents the trace of a $6\times6$ submatrix formed by selecting the middle region of the original matrix (the region of rows $4-9$ $\times$ column $4-9$).

The second one is the bulk energy gap under the condition PBCs, i.e., 
%%%
%%%
\begin{equation}
\begin{aligned}
\ln(\Delta E)=\ln(E_{N+1}-E_{N}),
\end{aligned}
\end{equation}
%%%
%%%
which can well exhibit the phase transition critical points in REPT process. Concretely speaking, for topological phase transition, the bulk gap closing and reopening will occur, whereas for localized phase transition, the corresponding bulk gap will close and not reopen. 

The third one is the key indicator of localization transition--the Lyapunov exponent. For the model of Eq.~\eqref{Hamiltonian}, the corresponding wave function of zero mode $\psi=\left\{\psi_{1,A}, \psi_{1,B}, \psi_{2,A}, \psi_{2,B}, \dots, \psi_{N,A}, \psi_{N,B}\right\}^{T}$, which can be solved by the Schrödinger equation $H\psi=0$. One can obtain the eigenequations $t_2\psi_{n,B}+t^{'}_{1,n+1}\psi_{n+1,B}=0$ and $t^{'}_{1,n}\psi_{n,A}+t_2\psi_{n+1,A}=0$. Then, the corresponding probability distribution of the zero mode wave function reads
%%%
%%%
\begin{equation}
\begin{aligned}
\psi_{n,A}=&(-1)^n \prod_{l=1}^n \frac{t_{1,l}^{'}}{t_2} \psi_{1, A}, \\
\psi_{n, B}=&(-1)^n \prod_{l=1}^n \frac{t_2}{t_{1,l+1}^{'}} \psi_{1, B}. 
\end{aligned}
\end{equation}
%%%
%%%
Then, one can obtain the Lyapunov exponent $\gamma$ of the zero mode for $N\to\infty$, which is the inverse of the localization length~\cite{san1,san2}, i.e.,
%%%
%%%
\begin{equation}
\gamma=\max \left\{\lim _{N \rightarrow \infty} \frac{1}{N} \ln \psi_{N, A}, \lim _{N \rightarrow \infty} \frac{1}{N} \ln \psi_{N, B}\right\}. \end{equation}
By set $\psi_{1, A}=\psi_{1, B}=1$, and by performing a straightforward calculation, one can obtain
%%%
%%%
\begin{equation}
\begin{aligned}
\gamma & =\lim _{N \rightarrow \infty} \frac{1}{N} \ln \psi_{N, A}=\lim _{N \rightarrow \infty} \frac{1}{N} \ln \psi_{N, B} \\
& =\left|\lim _{N \rightarrow \infty} \frac{1}{N} \sum_{l=1}^N\left(\ln |t_2|-\ln \left|t_{1,l}^{'}\right|\right)\right|.
\end{aligned}\label{E6}
\end{equation}
%%%
%%%
Generally, if the Lyapunov exponent $\gamma>0$ ($\gamma=0$), the corresponding wave function will have the characteristics of localization (extension). 

Since the generalized quasiperiodic modulation can be distinguished as bounded (Sec.~\ref{S3}) and unbounded (Sec.~\ref{S4}) cases, in the following sections, we will discuss these two typical cases, respectively.

\section{Topological REPT for the BOUNDED case}\label{S3}
\begin{figure}[bhtp]
\centering
\includegraphics[width=8.3cm]
{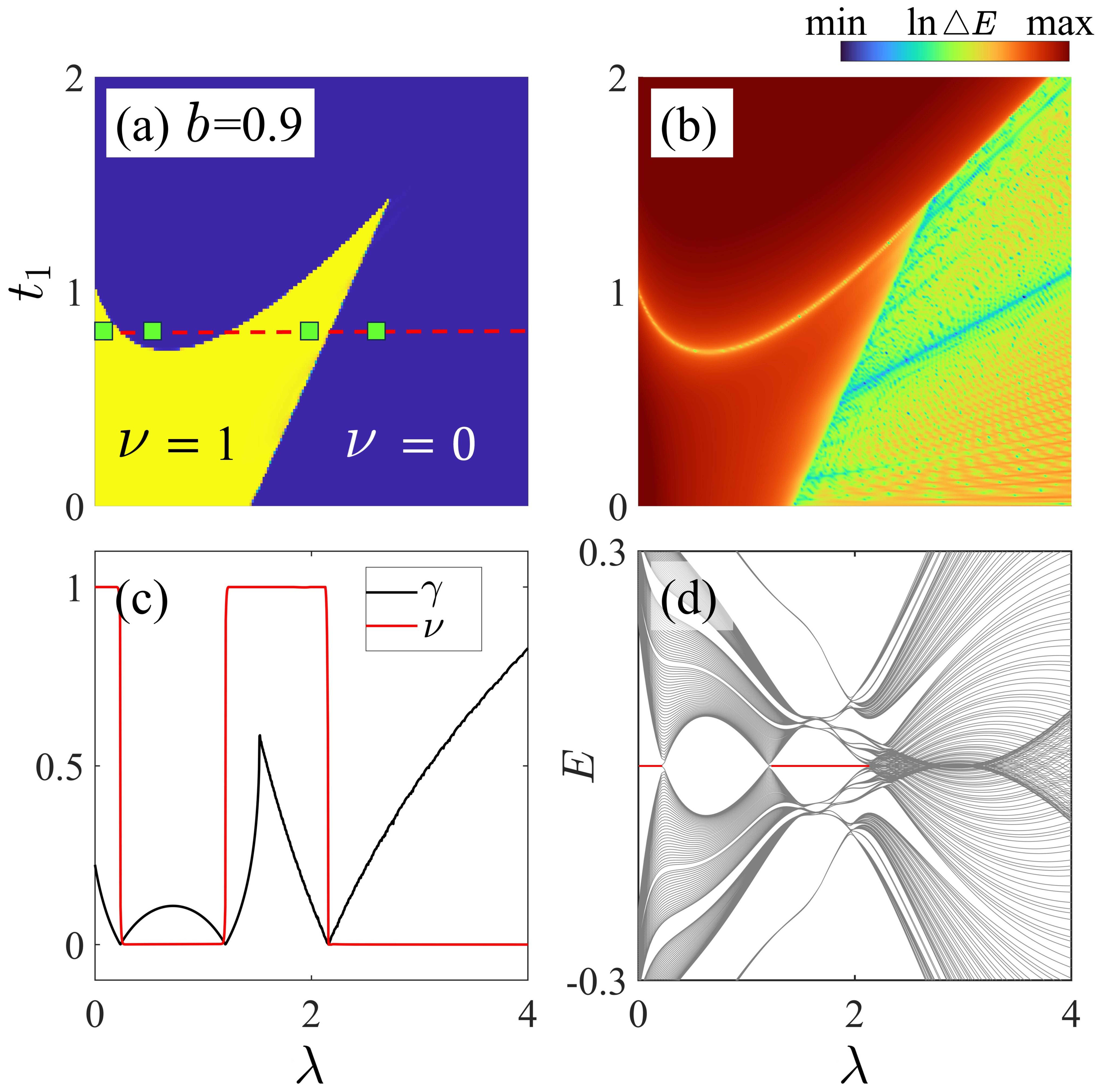}
\caption{(a) The real-space winding number $\nu$ as functions of $\lambda$ and $t_{1}$. The red dashed line corresponds to the line of $t_1=0.8$. (b) The bulk gap $(\Delta E)$ as functions of $\lambda$ and $t_1$. (c) The winding number (red solid line) and Lyapunov exponent (black solid line) versus quasiperiodic strength $\lambda$ with $t_1 = 0.8$. (d) The middle $200$ eigenenergies versus $\lambda$ with $t_1 = 0.8$. The emergence of topological zero modes are marked with red lines. Throughout, $b=0.9$.}
\label{2}
\end{figure}

Let's start with the bounded case, i.e., the case where $\left|b\right|<1$. Without loss of generality, we fix $b=0.9$ to show how the topological properties of the system change with $\lambda$ and $t_1$. The corresponding results are shown in Fig.~\ref{2}(a). We find that, unlike traditional topological phase transitions, topological REPT phenomena will occur in the range of parameter $t_1\in[0.7,1]$ for the model of Eq.~\eqref{Hamiltonian}. Taking $t_1=0.8$ (red dashed line) as an example. One can see that topological REPT phenomenon will emerge as the quasiperiodic strength $\lambda$ increases. Specifically, the system starts from a traditional TI phase. With the increase of $\lambda$, the system first enters the trivial phase, and then enters the TAI phase. Finally, when the disorder strength completely dominates, the system will enter the trivial region due to Anderson localization. We refer to this phenomenon of re-entrant from traditional TI into TAI as the type-I of topological REPT. 

%%%%%%
%%%%%%
%%%%%%
%%%%%% 

Besides, the bulk energy gap under periodic boundary conditions (PBCs), a quantity commonly used to indicate critical points, is used here to verify again the emergence of topological REPT. The corresponding gap is plotted in Fig.~\ref{2}(b). Here, we rescale the gap with $\ln$ function for a more intuitive display. In other words, the red region in the figure corresponds to the open bulk gap, while other colors all to the closed gap. As shown in the figure, the emergence of topological REPT is accompanied by the bulk gap's closing and reopening. Note that, the boundary in the upper right corner of Fig.~\ref{2}(b) indicates the traditional Anderson phase transition due to the increase in quasiperiodic strength. This also shows that bulk gap is universal as a critical point. In other words, it is not only limited to determining the critical points of topological phase transitions, but can also be used to determine the critical points of other various types of phase transitions. It is not difficult to find that for Anderson phase transition, the energy gap will remain unchanged after closing and will never be opened again.

Furthermore, in Fig.~\ref{2}(c), we show how the corresponding winding number and Lyapunov exponent of the system change with the $\lambda$. The result confirms once again that topological REPT can emerge in the system. It is worth noting that during topological REPT, the Lyapunov exponent always tends to be zero at all critical points of toplogical phase transitions (i.e., where the winding number changes abruptly). This is because topological protected edge states will always appear when the system is in a nontrivial phase (TI or TAI). Since these edge states are localized in the vicinty of boundary, the wave function also exhibits exponential decay, i.e., $\psi\propto e^{-\gamma n}$. On the other hand, with the increase of disorder, Anderson localization occurs in the system, and the corresponding wave function will be localized. The competition between the localized wave functions on the edge and in the middle of the atomic chain eventually causes the Lyapunov exponent corresponding to the critical point to approach zero. Fig.~\ref{2}(d) shows the energy spectra corresponding to the middle $200$ eigenvalues of the system. Similarly, through the appearance and disappearance of zero mode, one can find that topological REPT does emerge in the system.

\begin{figure}[bhtp]
\centering
\includegraphics[width=7.8cm]
{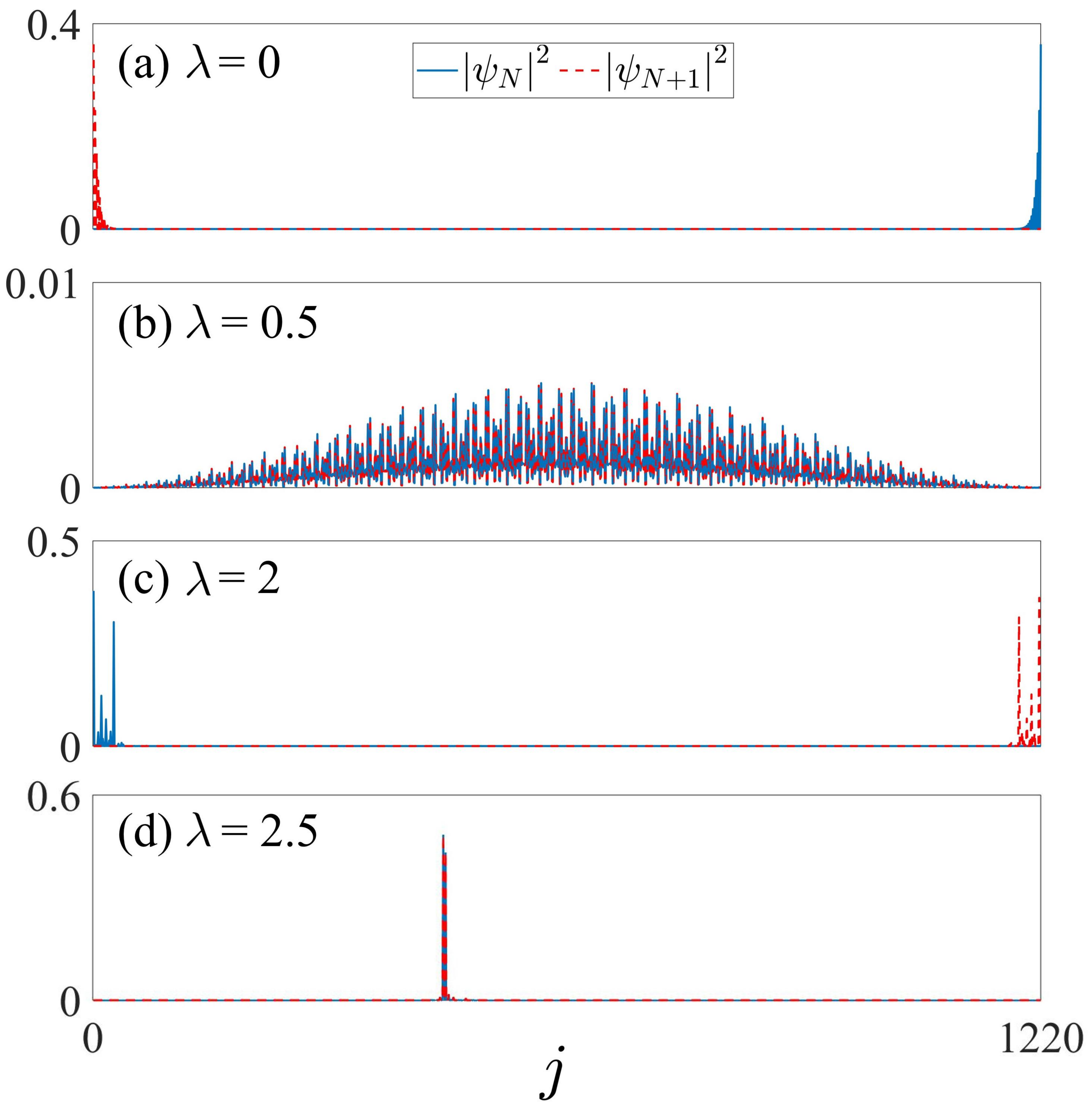}
\caption{Density distribution of the $N$-th and $N+1$-th eigenstates under the condition of $\lambda=0,~0.5,~2,~2.5$ (marked by green squares in Fig.~\ref{2}). Throughout, we set $t_1=0.8$.}
\label{3}
 \end{figure}

To show the topological REPT phenomenon more clearly, we plot the density distribution of the wave function corresponding to the topological zero-mode bands (the $610$-th and $611$-th eigenenergies) in Fig.~\ref{3}.

%%%%%%
%%%%%%
%%%%%%
%%%%%%

Fig.~\ref{3}(a)-(d) correspond to the parameter values of the green square in Fig.~\ref{2}(a), respectively. One can find that when the system is in the topological phase, the topological edge states will appear, while the edge states will not emerge in the trivial phase. Note that, because the reappearance of the topological or trivial phase are due to the localization properties caused by the strong quasiperiodic modulation, the corresponding wave function, which is of the localized state rather than the extended state [see Fig.~\ref{3}(d)], is therefore different from that of the first trivial region. With the visualized density distribution mentioned here, we reconfirm our conclusion that topological REPT does emerge in the system.
 
\section{TOPOLOGICAL REPT FOR THE
UNBOUNDED CASE}\label{S4}

\begin{figure}[bhtp]
\centering
\includegraphics[width=8.3cm]
{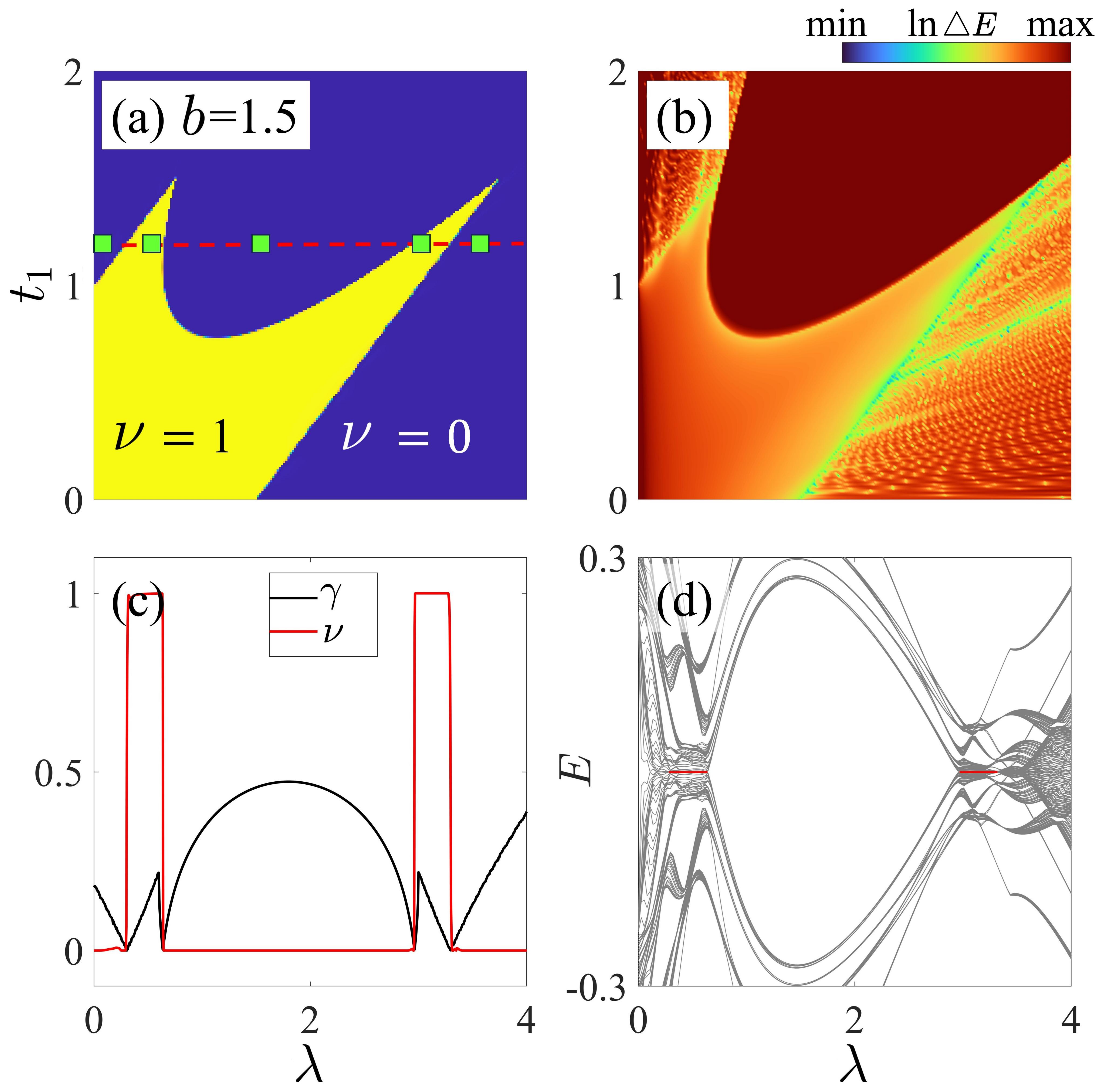}
\caption{(a) The real-space winding number $\nu$ as functions of $\lambda$ and $t_{1}$. The red dashed line corresponds to the line of $t_1=1.2$. (b) The bulk gap $(\Delta E)$ as functions of $\lambda$ and $t_1$. (c) The winding number (red solid line) and Lyapunov exponent (black solid line) versus quasiperiodic strength $\lambda$ with $t_1 = 1.2$. (d) The middle $200$ eigenenergies versus $\lambda$ with $t_1 = 1.2$. The emergence of topological zero modes are marked with red lines. Throughout, $b=1.5$.}
\label{4}
\end{figure}

Next, let's turn to the unbounded case, i.e., under the condition of $\left|b\right|\geq1$~\cite{TLiu2022,YCZhang2022}. Using winding number, we exhibit the corresponding topological phase diagram in the $t_1$-$\lambda$ plane (see in Fig.~\ref{4}). The results also exhibit that topological REPT emerges. In concrete terms, the system will first go from a trivial state to a TAI. Then, with the increase of quasiperiodic strength, the system will jump out of the TAI phase into a trivial phase. After that, it will enter the TAI phase again. Similarly, because the quasiperiodic strength will eventually dominate, the system will thus end up in a trivial localized phase. Note that, unlike type-I topological REPT (TI$\rightarrow$TAI), in the unbounded case, both topological phases of the system are TAI phases induced by quasiperiodic modulation. We name this type of REPT from TAI into TAI as type-II topological REPT. For the unbounded case, we also analyze the energy gap, winding number, Lyapunov exponent and topological zero module respectively, and the results consistently prove that topological REPT can occur in the system.

%%%%%%
%%%%%%
%%%%%%
%%%%%%

\begin{figure}[bhtp]
\centering
\includegraphics[width=7.8cm]
{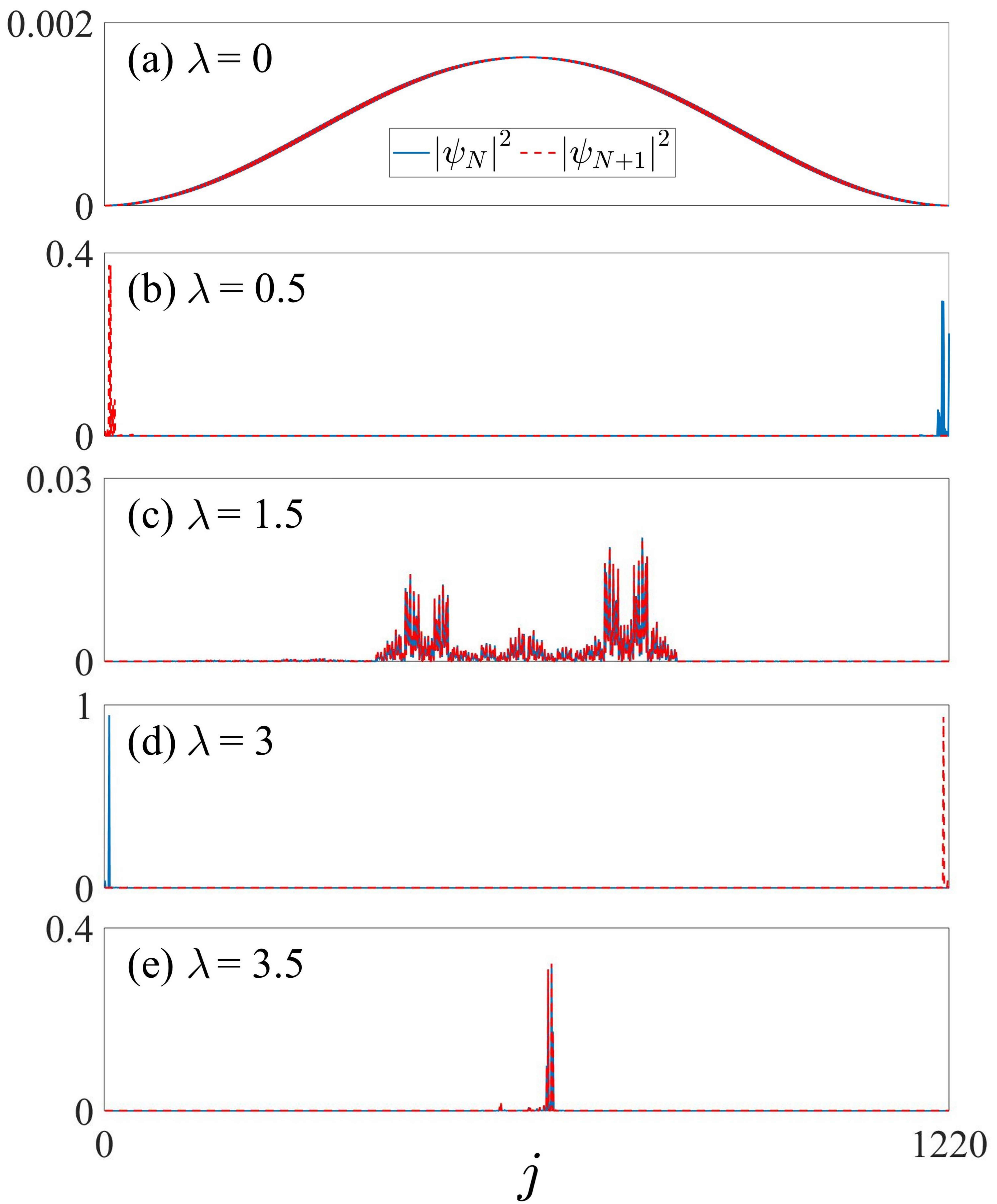}
\caption{Density distribution of the $N$-th and $N+1$-th eigenstates under the condition of $\lambda=0,~0.5,~1.5,~3,~3.5$ (marked by green squares in Fig.~\ref{4}). Throughout, we set $t_1=1.2$.}
\label{5}
\end{figure}

For a more visualized presentation, we calculate the corresponding eigenstate wave functions under different quasiperiodic strengths. Specifically, the system first shows the trivial phase not yet being localized. Then, with the increase of quasiperiodic intensity, we notice that the system begins to appear edge states, which is the evidence that the system has entered the topological phase. Subsequently, further increases of the quasiperiodic intensity pull the topological edge states back into the atomic chain, and this competition leads to the emergence of an intermediate state where the wave function structure is between the extended and localized states [see Fig.~\ref{5}(c)]. After that, the edge state will appear again, which indicates that the system again enters the topological phase. Finally, because the localization properties prevail, the wave function of the system shows characteristics of the localized state. These visualized results are consistent with the analysis of the core physical quantities in Fig.~\ref{5}, i.e., topological REPT emerges in the system.

%%%%%%
%%%%%%
%%%%%%
%%%%%%

\section{Discussions and CONCLUSIONS}
\label{S6}
 Topological REPT has recently received much attention because of its important theoretical and potential application value. Re-entrant integer quantum Hall effect, re-entrant quantum spin Hall effect, re-entrant quantum anomalous Hall effect, and re-entrant topological superconductivity have been successively discovered~\cite{z9,re51,z10,re52,513,514,zp11,zp12,zp13,516}. The systems involved are also diverse, including traditional topological systems~\cite{z9,re51,z10,re52,513,514}, topological superconductors~\cite{zp11,zp12,zp13,516}, hybrid bridging systems~\cite{zp14}, non-Hermitian systems~\cite{515}, and so on~\cite{517,pl1,518}. However, there are two differences between the topological re-entrant seen in these systems and the REPT phenomenon proposed in this paper, i.e., the inducers of REPT as well as the topological phases involved in REPT. For example, the reentrant integer quantum Hall effect occurs in a two-dimensional degenerate electron gas as the magnetic field changes~\cite{z9,re51,z10,re52}, and the bridging model of Kitaev chain and Haldane chain realize REPT from Majorana zero mode phase to Z2 topologically ordered phase through pairing interaction~\cite{zp14}. It can be seen clearly that the former is induced by magnetic fields, and the latter is induced by pairing interactions. While, both are fundamentally different from the phenomenon studied in this paper that is controlled by disorder. In addition, most of the previous studies involved traditional topological phases, while this work focuses on the REPT of the topological (Anderson) insulator phases in a weakly disordered region, and this non-traditional object of study constitutes the other major difference.

In summary, we introduce generalized quasiperiodic modulation into the primitive intracellular coupling of SSH model and thus obtain topological REPT. In addition, for bounded and unbounded cases, we give the phase diagram of the system by calculating winding number, and verify the diagram by Lyaponov exponent, zero mode and wave function properties. The results show that the topological REPT from TI to TAI occurs in the bounded case, while the topological phase transition from TAI to TAI occurs in the unbounded case. The SSH model discussed in this paper has recently been experimentally realized in Rydberg atomic array system~\cite{z3}. On this basis, we only need to use Rydberg single point manipulation technology to modulate the intracellular coupling intensity~\cite{r0,r1,r2,r3}, so as to realize the REPT phenomenon predicted in this paper. It is hoped that this paper can bring benefits to the research on topological characteristics of quasiperiodic systems, Rydberg atomic arrays and other related fields.

\section{ACKNOWLEDGEMENTS}
This work was supported by the National Key Research and Development Program of China (Grant No.2022YFA1405300), the National Natural Science Foundation of China (Grant No.12074180), the Guangdong Basic and Applied Basic Research Foundation (Grants No.2021A1515012350), and Open Fund of Key Laboratory of Atomic and Subatomic Structure and Quantum Control (Ministry of Education).

\appendix
\section{FINITE SIZE EFFECT}
\label{Sec.7}
\begin{figure}[bhtp]
\centering
\includegraphics[width=5.8cm]
{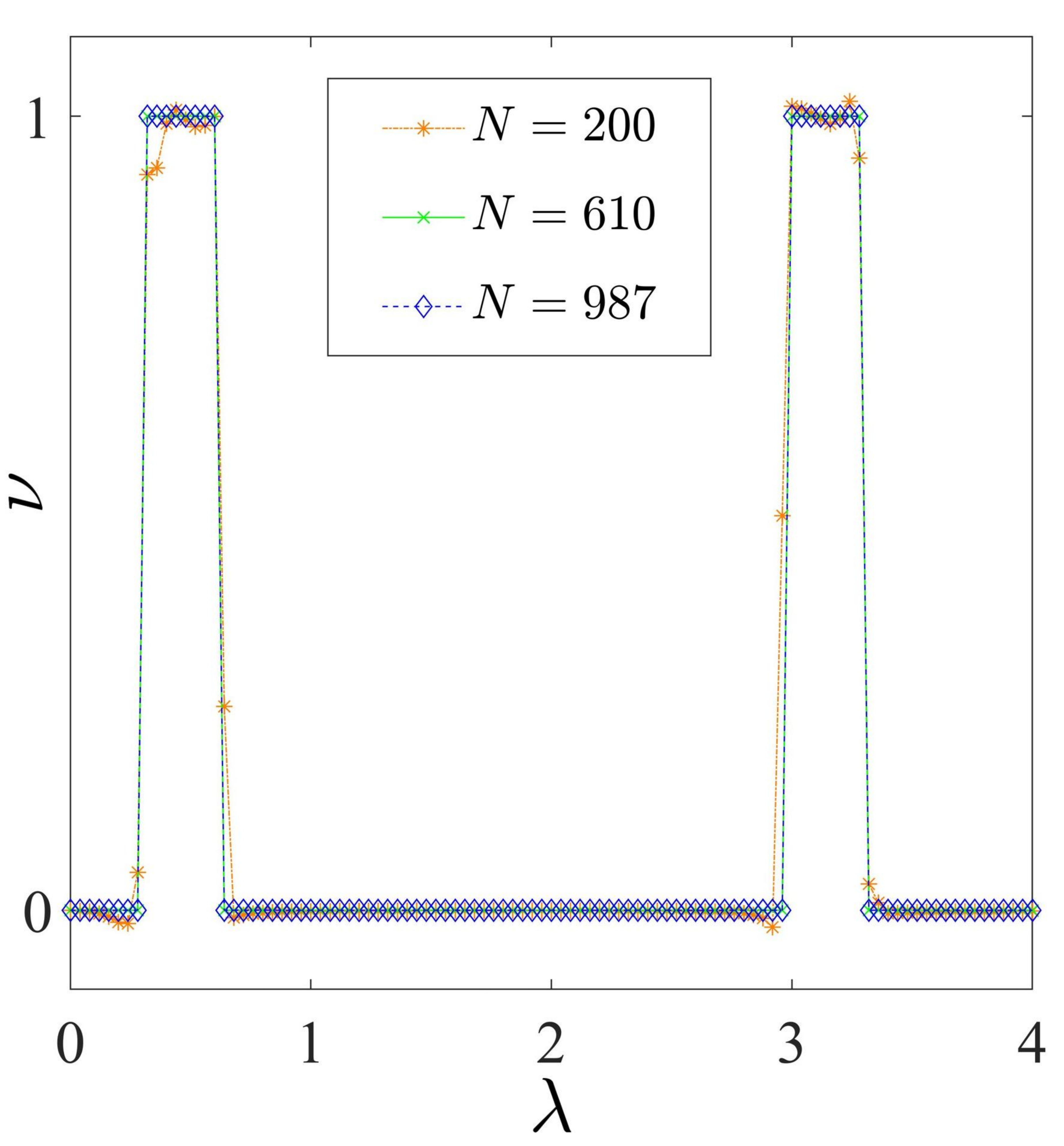}
\caption{Real-space winding number $\nu$ as functions of $\lambda$. The yellow dot, green solid, and blue dash lines are plotted to describe the system size with $N=200$, $N=610$, $N=987$, respectively. In all cases, $b=1.5$ and $t_{1}=1.2$.}
\label{6}
\end{figure}

In the numerical calculation, the total number of primitive cells we selected is $610$, i.e., the atomic chain contains $1220$ sites. It has been proved by many times of numerical calculation that, the size we chose can accurately show the characteristics of all the key quantities at a small computational cost. Taking the winding number as an example, we exhibit below the results of winding numbers with different sizes (see Fig.~\ref{6}).

%%%%%%
%%%%%%
%%%%%%
%%%%%%

From the figure, it can be seen clearly that When the system size is small, the calculated winding number is not correct. However, when the size becomes larger, one can calculate the corresponding winding number more accurately. After test, we find that $N=610$ is sufficient.

\begin{figure}[bhtp]
\centering
\includegraphics[width=8.3cm]
{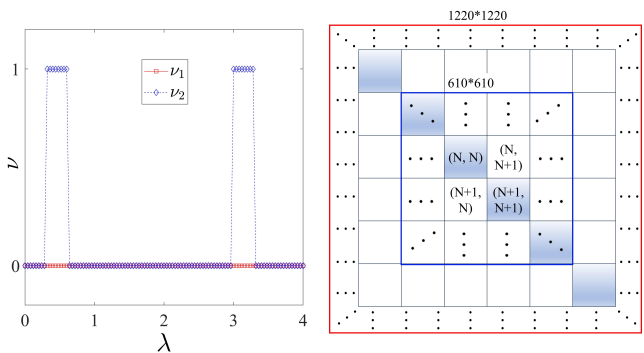}
\caption{The winding number for $N=610$ with different matrix traces. $\nu_{1}$ ($\nu_{2}$) is the average of trace over the whole (half) matrix. The corresponding trace region are marked by red (blue) square box. The colored boxes are color-matched to the line in the winding number plot.}
\label{7}
\end{figure}

On the other hand, due to the introduction of quasi-periodic modulation, the translational symmetry of the system is broken. Therefore, we have to study the topological properties through the real-space winding number~\cite{san1}. Since the calculation process is carried out for a finite size, the block effect of the matrix is revealed. The expression of winding number with different matrix block reads
%%%
%%%
\begin{equation}
\label{A1}
\nu_{i}=\frac{1}{L_i} \operatorname{Tr}_{i}(\Gamma \mathrm{Q}[\mathrm{Q}, \mathrm{X}]),
\end{equation}
%%%
%%%
where $\Gamma$, $\mathrm{Q}$, and $\mathrm{X}$ are matrices of $2N\times2N$. $\mathrm{Tr}_{i}$ denote traces for different matrix blocks. Due to the finite-size effect, one needs to trace a part of the matrix (see Fig.~\ref{7}). To demonstrate that region selection can indeed make a difference, we show winding numbers calculated by selecting different regions in Fig.~\ref{7}. 

%%%%%%
%%%%%%
%%%%%%
%%%%%%

With a fixed primitive size $N=610$, we show the case of trace with the whole matrix ($L$) and the half matrix blocks ($L/2$), respectively. It is not difficult to find that the winding number is best calculated using the trace of the central $L/2$ block. However, when we choose the trace of the whole matrix to calculate the winding number, there will be obvious errors in the result. This is because as a topology marker, the edge elements of the diagonal matrix in Eq.~\eqref{A1} lead to inaccurate results [see Fig.~\ref{7}(b)]. Therefore, as in previous studies ~\cite{001,traceb0,Fi1}, in main text, we also select the central $L/2$ region to calculate the winding number.

\section{TOPOLOGICAL PHASE DIAGRAM FOR THE BOUNDED TO UNBOUNDED CASE}\label{Sec.A}
\begin{figure}[bhtp]
\centering
\includegraphics[width=8.3cm]
{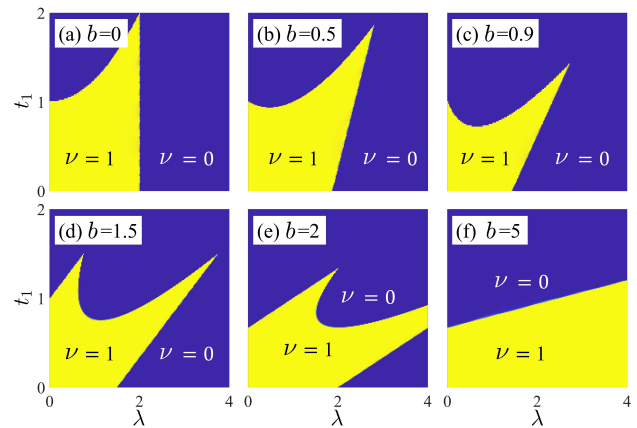}
\caption{(a), (b), and (c) correspond to bounded systems with b = 0, 0.5, and 0.9, respectively. (d), (e), and (f) correspond to unbounded systems with b = 1.5, 2, and 5, respectively.}
\label{8}
\end{figure}

The expression of the quasiperiodic modulation we introduce is,
%%%
%%%
\begin{equation}\label{eqA2}
{t_1^{'}}=t_1+\frac{\lambda \cos (2 \pi \alpha n+\theta)}{1-b \cos (2 \pi \alpha n+\theta)},
\end{equation}
%%%
%%%
One can find that the denominators of the quasiperiodic modulation term always converge for $b<1$, and diverge for $b\geq1$. These two different types of quasiperiodic modulation are often referred to as bounded and unbounded cases~\cite{YCZhang2022,si1}. As we all know, the change of the structure of quasiperiodic modulation will lead to changes in the corresponding topological properties, therefore there will be two different types of topological REPTs~\cite{tr1,tr2} (see Fig.~\ref{8}). The figure below shows how the phase diagram changes from bounded to unbounded case. It is not difficult to see that $b=1$ is the critical point from type-I topological REPT to type-II topological REPT. One can also see that the increase of the value of $b$ will cause the REPT region to move towards the larger $\lambda$ direction. Therefore, when the value of $b$ is fairly large, the REPT phenomenon cannot be seen in the relatively weak quasiperiodic region [see Fig.~\ref{8}(f)].

%%%%%%
%%%%%%

 \section{FIDELITY SUSCEPTIBILITY}\label{Sec.C}
\begin{figure}[bhtp]
\centering
\includegraphics[width=8.3cm]
{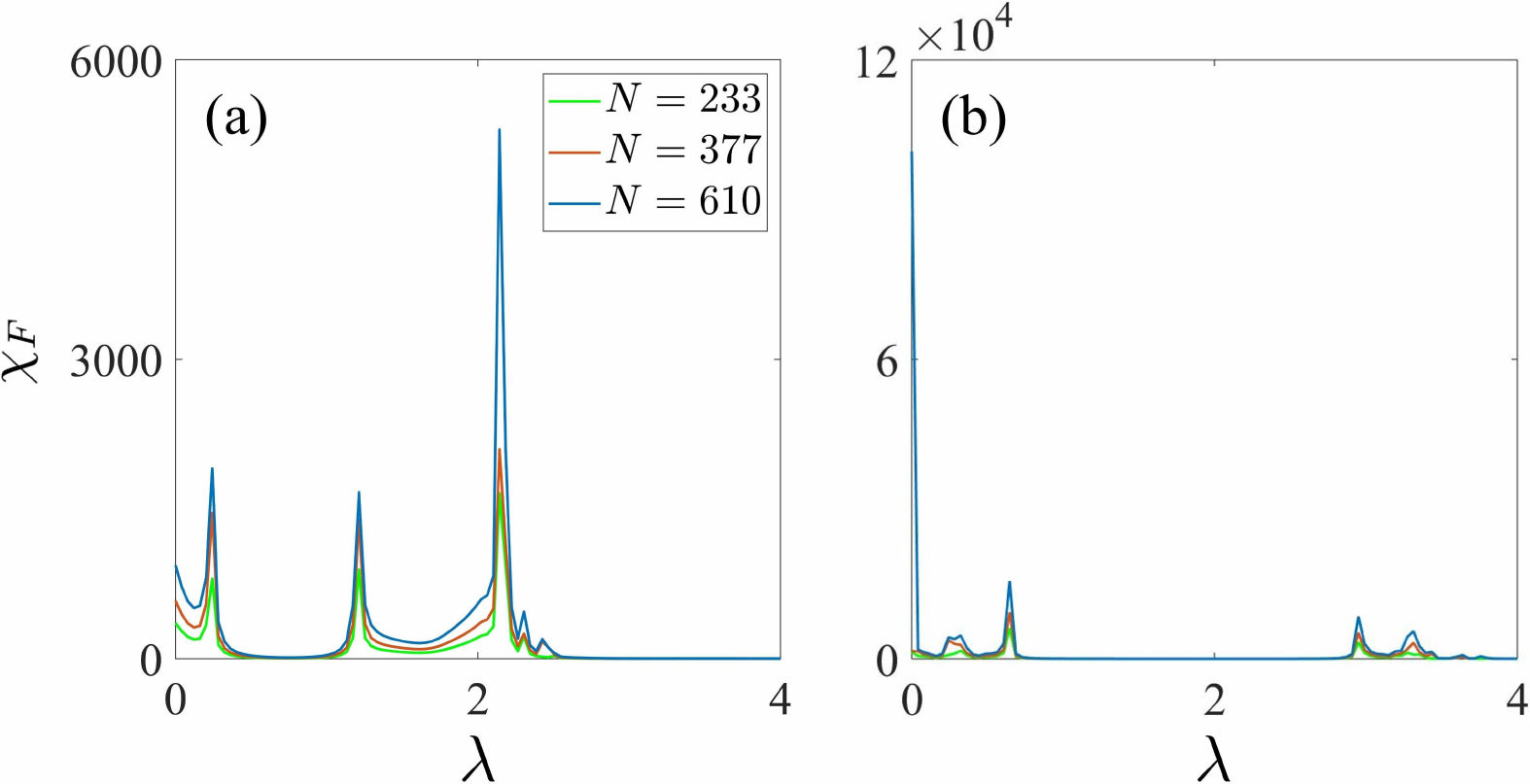}
\caption{ Fidelity susceptibility $\chi_{F}$ versus $\lambda$ with $b=0.9$ and $t_{1}=0.8$ (a), and $b=1.5$ and $t_{1}=1.2$ (b). System sizes are marked.}
\label{9}
\end{figure}

The fidelity susceptibility $\chi_{F}$ reads~\cite{FS1}
\begin{equation}
\label{eqA3}
\chi_{F}=-\frac{\partial^{2} F}{\partial(\delta \lambda)^{2}}=\lim _{\delta \lambda \rightarrow 0} \frac{-2 \log F}{(\delta \lambda)^{2}},
\end{equation}
with
\begin{equation}
\label{eqA4}
F\left(\Psi(\lambda), \Psi\left(\lambda^{\prime}\right)\right)=\left|\operatorname{det}\left(\boldsymbol{\Phi}_{\lambda}^{\dagger} \boldsymbol{\Phi}_{\lambda^{\prime}}\right)\right|,
\end{equation}
where $\Phi_{\lambda}$ is a $N\times M$ matrix containing the M filled single-particle eigenstates in its columns. We show how the fidelity changes with $\lambda$ for three different sizes. It is easy to find that the peak of the fidelity at critical points becomes sharper and sharper with the increase of the size, which indicates the appearance of ciritical points~(see Fig.~\ref{9}).

%%%%%%
%%%%%%


\begin{thebibliography}{99}
%anderson localization~\cite{PWAnderson1958}(21)
\bibitem{x1} J. Wang and S. C. Zhang, Topological states of condensed matter, Nat. Mater. {\bf 16}, 1062 (2017).
\bibitem{x2} M. He, H. Sun, and Q. L. He, Topological insulator: Spintronics and quantum computations, Front. Phys. {\bf 14}, 43401 (2019).
\bibitem{x3} A. Bansil, H. Lin, and T. Das, Colloquium: Topological band theory, Rev. Mod. Phys. {\bf 88}, 021004 (2016).
\bibitem{x4} H. Weng, X. Dai, and Z. Fang, Exploration and prediction of topological electronic materials based on first-principles calculations, MRS Bull. {\bf 39}, 849 (2014).
\bibitem{x5} P. Liu, J. R. Williams, and J. J. Cha, Topological nanomaterials, Nat. Rev. Mater. {\bf 4}, 479 (2019).
\bibitem{x6} D. Pesin and A. H. MacDonald, Spintronics and pseudospintronics in graphene and topological insulators, Nat. Mater. {\bf 11}, 409 (2012).
\bibitem{x7} D. Hsieh, D. Qian, L. Wray, Y. Xia, Y. S. Hor, R. J. Cava, and M. Z. Hasan, A topological Dirac insulator in a quantum spin Hall phase, Nature (London) {\bf 452}, 970 (2008).
\bibitem{x8} L. Fu and C. L. Kane, Topological insulators with inversion symmetry, Phys. Rev. B {\bf 76}, 045302 (2007).
\bibitem{x10} D. Hsieh, Y. Xia, D. Qian, L. Wray, J. H. Dil, F. Meier, J. Osterwalder, L. Patthey, J. G. Checkelsky, N. P. Ong, A. V. Fedorov, H. Lin, A. Bansil, D. Grauer, Y. S. Hor, R. J. Cava, and M. Z. Hasan, A tunable topological insulator in the spin helical Dirac transport regime, Nature (London) {\bf 460}, 1101 (2009).
\bibitem{x13} T. H. Hsieh, H. Lin, J. Liu, W. Duan, A. Bansil, and L. Fu, Topological crystalline insulators in the SnTe material class, Nat. Commun. {\bf 3}, 982 (2012).
\bibitem{x14} B. A. Bernevig, T. L. Hughes, and S.-C. Zhang, Quantum spin Hall effect and topological phase transition in HgTe quantum wells, Science {\bf 314}, 1757 (2006).
\bibitem{x15} Y. Li, P. Chen, G. Zhou, J. Li, J. Wu, B.-L. Gu, S. B. Zhang, and W. Duan, Dirac Fermions in Strongly Bound Graphene Systems, Phys. Rev. Lett. {\bf 109}, 206802 (2012).
\bibitem{x16} C. Weeks, J. Hu, J. Alicea, M. Franz, and R. Wu, Engineering a Robust Quantum Spin Hall State in Graphene via Adatom Deposition, Phys. Rev. X {\bf 1}, 021001 (2011).
\bibitem{x17} K.-H. Jin and S.-H. Jhi, Proximity-induced giant spin-orbit interaction in epitaxial graphene on a topological insulator, Phys. Rev. B {\bf 87}, 075442 (2013).
\bibitem{x18} Q.-X. Lv, Y.-X. Du, Z.-T. Liang, H.-Z. Liu, J.-H. Liang, L.-Q. Chen, L.-M. Zhou, S.-C. Zhang, D.-W. Zhang, B.-Q. Ai, H. Yan, and S.L. Zhu, Measurement of Spin Chern Numbers in Quantum Simulated Topological Insulators, Phys. Rev. Lett. {\bf 127}, 136802 (2021).
\bibitem{x19} F. Mei, Q. Guo, Y.-F. Yu, L. Xiao, S.-L. Zhu, and S. Jia, Digital Simulation of Topological Matter on Programmable Quantum Processors, Phys. Rev. Lett. {\bf 125}, 160503 (2020).
\bibitem{x20} X. Tan, D.-W. Zhang, Z. Yang, J. Chu, Y.-Q. Zhu, D. Li, X. Yang, S. Song, Z. Han, Z. Li, Y. Dong, H.-F. Yu, H. Yan, S.-L. Zhu, and Y. Yu, Experimental Measurement of the Quantum Metric Tensor and Related Topological Phase Transition with a Superconducting Qubit, Phys. Rev. Lett. {\bf 122}, 210401 (2019).

\bibitem{1} M. Z. Hasan and C. L. Kane, Colloquium: Topological insulators, Rev. Mod. Phys. {\bf 82}, 3045 (2010).
\bibitem{2} X.-L. Qi and S.-C. Zhang, Topological insulators and superconductors, Rev. Mod. Phys. {\bf 83}, 1057 (2011).
\bibitem{3} R. S. K. Mong and V. Shivamoggi, Edge states and the bulk boundary correspondence in Dirac Hamiltonians, Phys. Rev. B {\bf 83}, 125109 (2011).
\bibitem{4} K. Yatsugi, T. Yoshida, T. Mizoguchi, Y. Kuno, H. Iizuka, Y. Tadokoro, and Y. Hatsugai, Observation of bulk-edge correspondence in topological pumping based on a tunable electric circuit, Commun. Phys. {\bf 5}, 180 (2022).
\bibitem{5} S. Chen, L. Bu, C. Pan, C. Hou, F. Baronio, P. Grelu, and N. Akhmediev, Modulation instability–rogue wave correspondence hidden in integrable systems, Commun. Phys. {\bf 5}, 297 (2022).
\bibitem{6} Y. Hasegawa, Unifying speed limit, thermodynamic uncertainty relation and Heisenberg principle via bulk-boundary correspondence, Nat. Commun. {\bf 14}, 2828 (2023).
\bibitem{7} D.-W. Zhang, Y.-Q. Zhu, Y. X. Zhao, H. Yan, and S.-L. Zhu, Topological quantum matter with cold atoms, Adv. Phys. {\bf 67}, 253 (2019).
\bibitem{8} A. Nava, G. Campagnano, P. Sodano, and D. Giuliano, Lindblad master equation approach to the topological phase transition in the disordered Su-Schrieffer-Heeger model, Phys. Rev. B {\bf 107}, 035113 (2023).
\bibitem{9} E. G. Cinnirella, A. Nava, G. Campagnano, and D. Giuliano, Fate of high winding number topological phases in the disordered extended Su-Schrieffer-Heeger model, Phys. Rev. B {\bf 109}, 035114 (2024).
\bibitem{10} C.-A. Li, B. Fu, Z.-A. Hu, J. Li, and S.-Q. Shen, Topological Phase Transitions in Disordered Electric Quadrupole Insulators, Phys. Rev. Lett. {\bf 125}, 166801 (2020).

\bibitem{03} P. W. Anderson, Absence of diffusion in certain random lattices, Phys. Rev. {\bf 109}, 1492 (1958).
\bibitem{001} L.-Z. Tang, S.-N. Liu, G.-Q. Zhang, and D.-W. Zhang, Topological Anderson insulators with different bulk states in quasiperiodic chains, Phys. Rev. A {\bf 105}, 063327 (2022).
\bibitem{traceb0} Z. Lu, Z. Xu, and Y. Zhang, Exact mobility edges and topological Anderson insulating phase in a slowly varying quasiperiodic model, Ann. Phys. \textbf{534}, 2200203 (2022).
\bibitem{04} J. Li, R.-L. Chu, J. K. Jain, and S.-Q. Shen, Topological Anderson insulator, Phys. Rev. Lett. {\bf 102}, 136806 (2009).
\bibitem{05} C. W. Groth, M. Wimmer, A. R. Akhmerov, J. Tworzydło, and C. W. J. Beenakker, Theory of the topological anderson insulator, Phys. Rev. Lett. {\bf 103}, 196805 (2009).
\bibitem{06} A. Yamakage, K. Nomura, K.-I. Imura, and Y. Kuramoto, Disorder-induced multiple transition involving ${Z}_{2}$ topological insulator, J. Phys. Soc. Jpn. {\bf 80}, 053703 (2011). 
\bibitem{07} Y. Xing, L. Zhang, and J. Wang, Topological Anderson insulator phenomena, Phys. Rev. B {\bf 84}, 035110 (2011).

\bibitem{f6} H. Jiang, L. Wang, Q.-F. Sun, and X. C. Xie, Numerical study of the topological Anderson insulator in HgTe/CdTe quantum wells, Phys. Rev. B {\bf 80}, 165316 (2009).
\bibitem{f7} Y.-Y. Zhang, R.-L. Chu, F.-C. Zhang, and S.-Q. Shen, Localization and mobility gap in the topological Anderson insulator, Phys. Rev. B {\bf 85}, 035107 (2012).
\bibitem{f9} G.-Q. Zhang, L.-Z. Tang, L.-F. Zhang, D.-W. Zhang, and S. L. Zhu, Connecting topological Anderson and Mott insulators in disordered interacting fermionic systems, Phys. Rev. B {\bf 104}, L161118 (2021).
\bibitem{f10} S. N. Liu, G. Q. Zhang, L. Z. Tang, and D.-W. Zhang, Topological Anderson insulators induced by random binary disorders, Phys. Lett. A {\bf 104}, 128004 (2022).

\bibitem{h01} J. Song, H. Liu, H. Jiang, Q.-f. Sun, and X. Xie, Dependence of topological Anderson insulator on the type of disorder, Phys. Rev. B {\bf 85}, 195125 (2012).
\bibitem{h02} L. Chen, Q. Liu, X. Lin, X. Zhang, and X. Jiang, Disorder dependence of helical edge states in HgTe/CdTe quantum wells, New J. Phys. {\bf 14}, 043028 (2012).
\bibitem{h03} J. Song and E. Prodan, AIII and BDI topological systems at strong disorder, Phys. Rev. B {\bf 89}, 224203 (2014).
\bibitem{h04} H.-C. Hsu and T.-W. Chen, Topological Anderson insulating phases in the long-range Su-Schrieffer-Heeger model, Phys. Rev. B {\bf 102}, 205425 (2020).
\bibitem{h05} L. Lin, Y. Ke, and C. Lee, Real-space representation of the winding number for a one-dimensional chiral-symmetric topological insulator, Phys. Rev. B {\bf 103}, 224208 (2021).
\bibitem{h06} Z.-Q. Zhang, B.-L. Wu, J. Song, and H. Jiang, Topological Anderson insulator in electric circuits, Phys. Rev. B {\bf 100}, 184202 (2019).
\bibitem{h07} X. Shi, I. Kiorpelidis, R. Chaunsali, V. Achilleos, G. Theocharis, and J. Yang, Disorder-induced topological phase transition in a one-dimensional mechanical system, Phys. Rev. Res. {\bf 3}, 033012 (2021).
\bibitem{h08} D. Bagrets, K. W. Kim, S. Barkhofen, S. De, J. Sperling, C. Silberhorn, A. Altland, and T. Micklitz, Probing the topological Anderson transition with quantum walks, Phys. Rev. Res. {\bf 3}, 023183 (2021).
\bibitem{h1} S. Huang, Y.-Q. Zhu, Z. Li, Emergent non-Abelian Thouless pumping induced by the quasiperiodic disorder, Phys. Rev. A {\bf 109}, 052213 (2024).
\bibitem{h09} K. Roy, S. Roy, and S. Basu, Quasiperiodic disorder induced critical phases in a periodically driven dimerized p-wave Kitaev chain, Sci. Rep. {\bf 14}, 20603 (2024).
 
\bibitem{e1} X. Cui, R.-Y. Zhang, Z.-Q. Zhang, and C. T. Chan, Photonic $\mathbb{Z}_{2}$ Topological Anderson Insulators, Phys. Rev. Lett. {\bf 129}, 043902 (2022).
\bibitem{e2} X. Cheng, T. Qu, L. Xiao, S. Jia, J. Chen, and L. Zhang, Topological Anderson amorphous insulator, Phys. Rev. B {\bf 108}, L081110 (2023).
\bibitem{e3} W. Zhang, D. Zou, Q. Pei, W. He, J. Bao, H. Sun, and X. Zhang, Experimental Observation of Higher-Order Topological Anderson Insulators, Phys. Rev. Lett. {\bf 126}, 146802 (2021).
\bibitem{h2} Y.-B. Yang, K. Li, L.-M. Duan, and Y. Xu, Higher-order topological Anderson insulators, Phys. Rev. B {\bf 103}, 085408 (2021).
\bibitem{e4} Z.-W. Zuo, J.-R. Lin, and D. Kang, Topological inverse Anderson insulator, Phys. Rev. B {\bf 110}, 085157 (2024).
\bibitem{e5} R. Chen, X.-X. Yi, and B. Zhou, Four-dimensional topological Anderson insulator with an emergent second Chern number, Phys. Rev. B {\bf 108}, 085306 (2023).
\bibitem{e6} T. Peng, C.-B. Hua, R. Chen, D.-H. Xu, and B. Zhou, Topological Anderson insulators in an Ammann-Beenker quasicrystal and a snub-square crystal, Phys. Rev. B {\bf 103}, 085307 (2021).
\bibitem{e7} Y.-C, Zhang, Critical regions in a one-dimensional flat band lattice with a quasi-periodic potential, Sci Rep {\bf 14}, 17921 (2024). 

\bibitem{nh00} R. El-Ganainy, K. G. Makris, M. Khajavikhan, Z. H. Musslimani, S. Rotter, and D. N. Christodoulides, Non-Hermitian physics and PT symmetry, Nat. Phys. {\bf 14}, 11 (2018).
\bibitem{nh001} Y. Ashida, Z. Gong, and M. Ueda, Non-Hermitian Physics, Adv. Phys. 69, {\bf 3} (2020).
\bibitem{nh1} D.-W. Zhang, Y.-L. Chen, G.-Q. Zhang, L.-J. Lang, Z. Li, and S.-L. Zhu, Skin superfluid, topological Mott insulators, and asymmetric dynamics in an interacting non-Hermitian Aubry-Andr\'e-Harper model, Phys. Rev. B {\bf 101}, 235150 (2020).
\bibitem{nh002} E. J. Bergholtz, J. C. Budich, and F. K. Kunst, Exceptional topology of non-Hermitian systems, Rev. Mod. Phys. {\bf 93}, 015005 (2021).
\bibitem{nh01} N. Okuma and M. Sato, Non-Hermitian topological phenomena: A review, Annu. Rev. Condens. Matter Phys. 14, {\bf 83} (2023).
\bibitem{nh02} A. Li, H. Wei, M. Cotrufo, W. Chen, S. Mann, X. Ni, B. Xu, J. Chen, J. Wang, S. Fan, C.-W. Qiu, A. Al\'u, and L. Chen, Exceptional points and non-Hermitian photonics at the nanoscale, Nat. Nanotechnol. {\bf 18}, 706 (2023).
\bibitem{nh03} R. Lin, T. Tai, L. Li, and C.H. Lee, Topological non-Hermitian skin effect, Front. Phys. {\bf 18}, 53605 (2023).
\bibitem{nh04} D. Halder, S. Ganguly, and S. Basu, Properties of the non-Hermitian SSH model: Role of PT symmetry, J. Phys.: Condens. Matter {\bf 35}, 105901 (2023).
\bibitem{nh05} D. Halder, R. Thomale, and S. Basu, Circuit realization of a two-orbital non-hermitian tight-binding chain, Phys. Rev. B {\bf 109}, 115407 (2024).


\bibitem{z1} D.-W Zhang, L. Z. Tang, L. J. Lang, H. Yan, and S. L. Zhu, Non-Hermitian topological Anderson insulators, Sci. China-Phys. Mech. Astron. {\bf 63}, 267062 (2020).%非厄密
\bibitem{z02} Q. Lin, T. Li, L. Xiao, K. Wang, W. Yi, and P. Xue, Observation of non-Hermitian topological Anderson insulator in quantum dynamics, Nat. Commun. {\bf 13}, 3229 (2022).
\bibitem{z03} H. F. Liu, Z. X. Su, Z. Q. Zhang, and H. Jiang, Topological Anderson insulator in two-dimensional non-Hermitian systems, Chin. Phys. B {\bf 29}, 050502 (2020).
\bibitem{z04} L. Z. Tang, L. F. Zhang, G. Q. Zhang, and D.-W. Zhang, Topological Anderson insulators in two-dimensional non-Hermitian disordered systems, Phys. Rev. A {\bf 101}, 063612 (2020).

\bibitem{z2} L. B. Shao, S. L. Zhu, L. Sheng, D. Y. Xing, and Z. D. Wang, Realizing and Detecting the Quantum Hall Effect without Landau Levels by Using Ultracold Atoms, Phys. Rev. Lett. {\bf 101}, 246810 (2008).
\bibitem{z3} E. J. Meier, F. A. An, A. Dauphin, M. Maffei, P. Massignan, T. L. Hughes, and B. Gadway, Observation of the topological Anderson insulator in disordered atomic wires, Science {\bf 362}, 929 (2018).
\bibitem{zz0} D.-W. Zhang, Y.-Q. Zhu, Y.X. Zhao, H. Yan, and S.-L. Zhu, Topological quantum matter with cold atoms, Adv. Phys. {\bf 67}, 253 (2018).
\bibitem{z4} S. St\"utzer, Y. Plotnik, Y. Lumer, P. Titum, N. H. Lindner, M. Segev, M.C. Rechtsman, and A. Szameit, Photonic topological Anderson insulators, Nature (London) {\bf 560}, 461 (2018).%%%%%%%%%%2

\bibitem{z5} G.-G. Liu, Y. Yang, X. Ren, H. Xue, X. Lin, Y.-H. Hu, H. X. Sun, B. Peng, P. Zhou, Y. Chong, and B. Zhang, Topological Anderson Insulator in Disordered Photonic Crystals, Phys. Rev. Lett. {\bf 125}, 133603 (2020).
\bibitem{z6} F. Zangeneh-Nejad and R. Fleury, Disorder-induced signal filtering with topological metamaterials, Adv. Mater. {\bf 32}, 2001034 (2020).
\bibitem{002} X. Li, H. Xu, J. Wang, L.-Z. Tang, D.-W. Zhang, C. Yang, T. Su, C. Wang, Z. Mi, W. S, X. Liang, M. Chen, C. Li, Y. Zhang, K. Linghu, J. Han, W. Liu, Y. Feng, P. Liu, G. Xue, J. Zhang, Y. Jin, S.-L. Zhu, H. Yu, S. P. Zhao, and Q.-K. Xue, Mapping the topology-localization phase diagram with quasiperiodic disorder using a programmable superconducting simulator, Phys. Rev. Res. {\bf 6}, L042038 (2024).
\bibitem{z7} W. Zhang, D. Zou, Q. Pei, W. He, J. Bao, H. J. Sun, and X. Zhang, Experimental Observation of Higher-Order Topological Anderson Insulators, Phys. Rev. Lett. {\bf 126}, 146802 (2021).
\bibitem{z8} H. Fujii, T. Okamoto, T. Shigeoka, and N. Iwata, Reentrant ferromagnetism observed in SmMn2Ge2, Solid State Commun. {\bf 53}, 715 (1985).
\bibitem{a0} S. Aubry and G. Andr\'e, Analyticity breaking and Anderson localization in incommensurate lattices, Ann. Israel Phys. Soc {\bf 3}, 18 (1980).
\bibitem{a1} S. Roy, T. Mishra, B. Tanatar, and S. Basu, Reentrant Localization Transition in a Quasiperiodic Chain, Phys. Rev. Lett. {\bf 126}, 106803 (2021).

\bibitem{519} Z.-W. Zuo and D. Kang, Reentrant localization transition in the Su-Schrieffer-Heeger model with random-dimer disorder, Phys. Rev. A {\bf 106}, 013305 (2022).
\bibitem{512} Z.-S. Xu, J. Gao, A. Iovan, I. M. Khaymovich, V. Zwiller, and A. W. Elshaari, Observation of reentrant metal-insulator transition in a random-dimer disordered SSH lattice, npj Nanophoton {\bf 1}, 8 (2024).
\bibitem{520} W. Han and L. Zhou, Dimerization-induced mobility edges and multiple reentrant localization transitions in non-Hermitian quasicrystals, Phys. Rev. B {\bf 105}, 054204 (2022).
\bibitem{521} H. Wang, X. Zheng, J. Chen, L. Xiao, S. Jia, and L. Zhang, Fate of the reentrant localization phenomenon in the one-dimensional dimerized quasiperiodic chain with long-range hopping, Phys. Rev. B {\bf 107}, 075128 (2023).
\bibitem{tr1} M. Tezuka and N. Kawakami, Reentrant topological transitions in a quantum wire/superconductor system with quasiperiodic lattice modulation, Phys. Rev. B {\bf 85}, 140508(R) (2012).

\bibitem{tr2} Z. Lu, Y. Zhang, and Z. Xu, Reentrant Localization Transitions in a Topological Anderson Insulator: A Study of a Generalized Su-Schrieffer-Heeger Quasicrystal, Front. Phys. {\bf 20}, 024204 (2025).

\bibitem{san0} S. Ganeshan, J. H. Pixley, and S. Das Sarma, Nearest Neighbor Tight Binding Models with an Exact Mobility Edge in One Dimension, Phys. Rev. Lett. {\bf 114}, 146601 (2015).

\bibitem{ssh1} W. P. Su, J. R. Schrieffer, and A. J. Heeger, Soliton excitations in polyacetylene, Phys. Rev. B {\bf 22}, 2099 (1980).

\bibitem{san1} I. Mondragon-Shem, T. L. Hughes, J. Song, and E. Prodan, Topological Criticality in the Chiral-Symmetric AIII Class at Strong Disorder, Phys. Rev. Lett. {\bf 113}, 046802 (2014).
\bibitem{san2} J. A. Scales and E. S. Van Vleck, Lyapunov exponents and localization in randomly layered media, J. Comput. Phys. {\bf 133}, 27 (1997).

\bibitem{TLiu2022} T. Liu, X. Xia, S. Longhi, and L. Sanchez-Palencia, Anomalous mobility edges in one-dimensional quasiperiodic models, SciPost Phys. {\bf 12}, 27 (2022).
\bibitem{YCZhang2022} Y.-C. Zhang and Y.-Y. Zhang, Lyapunov exponent, mobility edges, and critical region in the generalized Aubry-Andr\'e model with an unbounded quasiperiodic potential, Phys. Rev. B {\bf 105}, 174206 (2022).
\bibitem{z9} J. P. Eisenstein, K. B. Cooper, L. N. Pfeiffer, and K. W. West, Insulating and Fractional Quantum Hall States in the First Excited Landau Level, Phys. Rev. Lett. {\bf 88}, 076801 (2002).
\bibitem{re51} M. O. Goerbig, P. Lederer, and C. M. Smith, Microscopic theory of the reentrant integer quantum Hall effect in the first and second excited Landau levels, Phys. Rev. B {\bf 68}, 241302 (2003).
\bibitem{z10} A. Kumar, G. A. Csáthy, M. J. Manfra, L. N. Pfeiffer, and K. W. West, Nonconventional Odd-Denominator Fractional Quantum Hall States in the Second Landau Level, Phys. Rev. Lett. {\bf 105}, 246808 (2010).
\bibitem{re52} Y. Liu, C. G. Pappas, M. Shayegan, L. N. Pfeiffer, K. W. West, and K. W. Baldwin, Observation of Reentrant Integer Quantum Hall States in the Lowest Landau Level, Phys. Rev. Lett. {\bf 109}, 036801 (2012).
\bibitem{513} X. Cheng, X. Zheng, J. Chen, L. Xiao, S. Jia, L. Zhang, Reentrant quantum spin Hall effect in the presence of an exchange field, Phys. Rev. B {\bf 107} (2023) 035117.
\bibitem{514} C.-Z. Chen, J. Qi, D.-H. Xu, and X. C. Xie, Evolution of Berry curvature and reentrant quantum anomalous Hall effect in an intrinsic magnetic topological insulator, Sci. China, Phys. Mech. Astron. {\bf 64}, 127211 (2021).
\bibitem{zp11} M. Tezuka and N. Kawakami, Reentrant topological transitions in a quantum wire/superconductor system with quasiperiodic lattice modulation, Phys. Rev. B {\bf 85}, 140508 (2012).
\bibitem{zp12} E. Lucioni, B. Deissler, L. Tanzi, G. Roati, M. Zaccanti, M. Modugno, M. Larcher, F. Dalfovo, M. Inguscio, and G. Modugno, Observation of Subdiffusion in a Disordered Interacting System, Phys. Rev. Lett. {\bf 106}, 230403 (2011).
\bibitem{zp13} M. Tezuka and N. Kawakami, Reentrant topological transitions with Majorana end states in one-dimensional superconductors by lattice modulation, Phys. Rev. B {\bf 88}, 155428 (2013).
\bibitem{516} M.-T. Rieder, P. W. Brouwer, and I. Adagideli, Reentrant topological phase transitions in a disordered spinless superconducting wire, Phys. Rev. B {\bf 88}, 060509 (2013).



\bibitem{zp14} T. Sugimoto, M. Ohtsu, and T. Tohyama, Reentrant topological phase transition in a bridging model between Kitaev and Haldane chains, Phys. Rev. B {\bf 96}, 245118 (2017).
\bibitem{515} A. Padhan, S. R. Padhi, and T. Mishra, Complete delocalization and reentrant topological transition in a non-Hermitian quasiperiodic lattice, Phys. Rev. B {\bf 109}, L020203 (2024).


\bibitem{517} M. Zhang, D. C. Zou and R. H. Yue, Reentrant phase transitions and triple points of topological AdS black holes in Born-Infeld-massive gravity, Adv. High Energy Phys. {\bf 2017}, 3819246 (2017).
%
\bibitem{pl1} W. Beugeling, C. X. Liu, E. G. Novik, L. W. Molenkamp, and C. M. Smith, Reentrant topological phases in Mn-doped HgTe quantum wells, Phys. Rev. B {\bf 85}, 195304 (2012). 


\bibitem{518} N. P. Mitchell, A. M. Turner, and W. T. Irvine, Real-space origin of topological band gaps, localization, and reentrant phase transitions in gyroscopic metamaterials, Physical Review E {\bf 104}, 025007 (2021).










\bibitem{r0} G. Semeghini, H. Levine, A. Keesling, S. Ebadi, T. T. Wang, D. Bluvstein, R. Verresen, H. Pichler, M. Kalinowski, R. Samajdar \textit{et al}., Probing topological spin liquids on a programmable quantum simulator, Science {\bf 374}, 1242 (2021).
\bibitem{r1} D. Bluvstein, H. Levine, G. Semeghini, T. T. Wang, S. Ebadi, M. Kalinowski, A. Keesling, N. Maskara, H. Pichler, M. Greiner, V. Vuleti\'c, and M. D. Lukin, A quantum processor based on coherent transport of entangled atom arrays, Nature (London) {\bf 604}, 451 (2022).
\bibitem{r2} D. Bluvstein, S. J. Evered, A. A. Geim, S. H. Li, H. Zhou, T. Manovitz, S. Ebadi, M. Cain, M. Kalinowski, D. Hangleiter, J. P. B. Ataides, N. Maskara, I. Cong, X. Gao, P. S. Rodriguez, T. Karolyshyn, G. Semeghini, M. J. Gullans, M. Greiner, V. Vuleti\'c, and M. D. Lukin, Logical quantum processor based on reconfigurable atom arrays, Nature (London) {\bf 626}, 58 (2024).
\bibitem{r3} T. Manovitz, S. H. Li, S. Ebadi, R. Samajdar, A. A. Geim, S. J. Evered, D. Bluvstein, H. Zhou, N. U. Koyluoglu, J. Feldmeier, P. E. Dolgirev, N. Maskara, M. Kalinowski, S. Sachdev, D. A. Huse, M. Greiner, V. Vuleti\'c, M. D. Lukin, Quantum coarsening and collective dynamics on a programmable quantum simulator, arXiv: 2407. 03249.

\bibitem{Fi1} G. Jin, D. O. Oriekhov, L. J. Splitthoff, and E. Greplova, Topological finite size effect in one-dimensional chiral symmetric systems, arXiv:2411.17822.
\bibitem{si1} B. Simon and T. Spencer, Trace class perturbations and the absence of absolutely continuous spectra, Commun. Math. Phys. {\bf 125}, 113 (1989).
\bibitem{FS1} M. Gonçalves, Entanglement entropy scaling in critical phases of one-dimensional quasiperiodic systems, Phys. Rev. B {\bf 109}, 104202 (2024).




\end{thebibliography}
\end{document}